\documentclass[journal,draftcls,onecolumn,12pt,twoside]{IEEEtran}

\usepackage[T1]{fontenc}
\usepackage{amsmath}
\usepackage{pdfpages}
\usepackage{mathrsfs}

\usepackage{hyperref}

\usepackage[cmintegrals]{newtxmath}

\newtheorem{theorem}{Theorem}

\newtheorem{corollary}{Corollary}
\newtheorem{definition}{Definition}
\newtheorem{example}{Example}
\newtheorem{lemma}{Lemma}

\newtheorem{proposition}{Proposition}

\begin{document}

\title{Canonical form of linear subspaces and coding invariants: the poset metric point of view}

\author{Jerry~Anderson~Pinheiro and Marcelo~Firer%
\thanks{The authors are  at Institute of Mathematics, Statistics and Scientific Computing, University of Campinas, Brazil, e-mail: jerryapinheiro@gmail.com, mfirer@ime.unicamp.br.}
}

\maketitle

\begin{abstract}
In this work we introduce the concept of a sub-space decomposition, subject to a partition of the coordinates. Considering metrics determined by partial orders in the set of coordinates, the so called poset metrics, we show the existence of maximal decompositions according to the metric. These decompositions turns to be an important tool to obtain the canonical form for codes over any poset metrics and to obtain bounds for important invariants such as the packing radius of a linear subspace. Furthermore, using maximal decompositions, we are able to reduce and optimize the full lookup table algorithm for the syndrome decoding process.
\end{abstract}


\IEEEpeerreviewmaketitle

\tableofcontents

\section{Introduction}


Coding theory is a mathematical theory that deals with many different aspects that emerges in the engineering of communication. A significant and classical part of it, the study of block-codes, is a subject that deals with many geometric and computational aspects of linear subspaces of finite dimensional vector spaces over finite fields.   

The most classic metric considered in this context is a metric that counts the number of different coordinates between two vectors, and it is named after R. W. Hamming, which introduced it in a seminal work \cite[ 1950]{ham}.  Later, in (\cite{ul}, 1957) and (\cite{lee}, 1958), another well-known metric was proposed, the Lee metric.

Up to our knowledge, the first work considering metrics in a general approach is a short communication of S. W. Golomb (\cite{golomb1969general}, 1969). In that work, Golomb described a family of additive metrics (metrics defined over an alphabet and additively extended to a set of words with a fixed length) which are still being investigated nowadays, as we can see in \cite{qureshiperfect}. Recently, due to the development of new communication channels and models, new families of metrics has been studied in the context of coding theory, as can be seen, for example, in \cite{brualdi1995codes}, \cite{alves2008error}, \cite{gab} and \cite{deza2009encyclopedia}.

The family of \textit{poset metrics} was defined by Brualdi, Graves and Lawrence, in (\cite{brualdi1995codes}, 1995) as an extension of the metrics proposed by Niederreiter in \cite{niederreiter1991combinatorial} and \cite{niederreiter1992orthogonal}. This is a large family of metrics which are determined by partial orders and generalizes the classical Hamming metric. Since then, it has been studied intensively, in order to understand the behaviour of metric invariants that are relevant to coding theory. 

As it frequently happens with generalizations, besides the possible inherent merits, some side effects may arise, either by giving a more profound understanding about the nature of previously established knowledge or by posing some new interesting (and difficult) questions. This is the case of poset metrics.

Postponing the details for the next section, a poset metric depends on the non-zero positions of a vector and also on the level of these positions in a partial order (or the level in the Hasse diagram). A significant part of the knowledge available for poset metrics can be understood once we consider a decomposition of a code (a vector subspace) which depend on the levels of the poset.  

Considering the metrics determined by the family of the so-called hierarchical posets, one can fully grasp the role of such decompositions. Being studied since the 90's, results previously proved\footnote{Results concerning the MacWilliams identity  \cite{classification:macwilliamsidentity}, association schemes \cite{classification:schemes}, extension of isometries  \cite{extensiontheoremforfiniterings3} (MacWilliams Extension Theorem)  and the relation between the packing radius and the minimum distance \cite{felix2012canonical}.} by many researchers were restated in an alternative and simple way \cite{MachadoPinheiroFirer}, all the proofs were derived from the existence of a canonical decomposition.



Besides the canonical-systematic form introduced in \cite{felix2012canonical} for hierarchical poset metrics, the only known attempt to construct standard forms for poset metrics was made in \cite{alves2011standard}, where a standard form for a particular case (Niederreiter-Rosenbloom-Tsfasman (NRT), or metrics induced by orders consisting of multiple disjoint chains with same cardinality) is presented. In that work, one can see that the standard form is not unique (in any possible sense). In a matter of fact,  unicity of such a pre-established form is a characteristic of hierarchical posets \cite[Theorem 3]{MachadoPinheiroFirer}. 

Since unicity of the form is not available, we strives for the best possible option:  canonical forms determined by maximal decompositions, whose unicity is obtained according to the degree of the decomposition. Hence, the main subject of this work is how to decompose a  subspace ``respecting'' the metric structure. For the reader that is not particularly interested in coding theory, despite the use of a terminology from coding theory, there is essentially a very short lexicon that needs to be translated: instead of ``code'' and ``$[n,k]_q$''-code, one should read ``vector subspace'' and ``$k$-dimensional vectors subspace of $\mathbb{F}_q^n$'', respectively; instead of ``$\mathcal{C}\subseteq\mathbb{F}_q^n$'', simply writes ``$V\subseteq\mathbb{F}_q^n$''.


This work is organized as follows. In Section \ref{secposet} we introduce the basic concepts of this work (poset metrics and $P$-decomposition) and describe an algorithm for decomposing a code.  In Section \ref{sec:refinement} we show that the decomposition of a subspace is ``well behaved'' in the sense that when different posets are comparable, so are the corresponding decompositions. Using this fact and the knowledge about hierarchical poset metrics, in part \ref{sub:hierar_bouds} we produce bounds for the degree of a maximal $P$-decomposition of a code, where $P$ is an arbitrary poset. Finally, on Section \ref{sec:applications} we go back to the coding ambient and give some applications: the determination of the packing radius (part \ref{sub:pack}) and the possible simplification of syndrome decoding algorithm  (part \ref{sub:syndrome}).


\section{Poset Metrics and Decomposition of Subspaces}\label{secposet}

This section has three parts. In the first part we summarize the basic concepts about poset metrics. In the second we introduce a new concept, object of study of this work: how different poset metrics allow different decompositions of a subspace. After doing so, we introduce the maximal and the primary $P$-decompositions of a subspace. The third part is devoted to an algorithm to find a maximal $P$-decompositions.


\subsection{Posets and Metrics}  \label{sub:poset}

Let $P=([n],\leqslant_P)$ be a partially ordered set (or poset), where  $[n]:=\{1,\ldots,n\}$ and $\leqslant_P$ is an order relation. We may denote the order relation only as $a\leqslant b$ if the poset is immaterial or clear in the context. A set $I\subset [n]$ is an \textit{order ideal} of $P$ if for every $x\in I$ and $y\in[n]$ with $y\leqslant_P x$ implies $y\in I$. Given  $E \subseteq [n]$, the smallest ideal  of $P$ containing $E$ is denoted by $\langle E \rangle_P$ and called the \textit{ideal generated by} $E$. The maximal elements (according to $P$) of $E$ will be denoted by $\mathcal{M}(E)$. In case $E=\{a\}$ is a singleton, we denote $\langle E \rangle_P=\langle a \rangle_P$.

An \emph{order isomorphism}  $f:P\rightarrow Q$ between posets $P$ and $Q$ over $[n]$ is a bijection $f:[n]\rightarrow [n]$ such that $a\leqslant_P b$ if, and only, if $f(a)\leqslant_Q f(b)$, for all  $a,b\in [n]$.   In the case $P=Q$, $f$ is said to be a $P$-automorphism. The group of all automorphisms of $P$ will be denoted by $Aut(P)$.

Given a poset  $P=([n],\leqslant_P)$ and a subset $X\subset [n]$, considering the order on $X$ inherited from $P$, we say that $(X,\leqslant_P) $ is a \emph{sub-poset} of $P$ and denote $X\subseteq P$.
A \emph{chain} is a poset that is totally ordered, that is, for any $a,b\in [n]$, either $a\leqslant b$ or $b\leqslant a$. An \emph{anti-chain} is a poset with no relations excepts the trivial relations $a\leqslant a$.

Given $a\in P$, the \emph{height} of $a$ according to $P$ is the maximal length of a chain in $P$ that has $a$ as a maximal element, i.e.,
\[
	h(a)=\max \{|C| \ : \ C\subseteq P \text{ is a chain }  \text{ and } b\leqslant_P a \ \forall \ b\in C\},
\] 
and the \emph{height of $P$} is $h(P)=\max \{h(a):a\in[n]\}$.
The $i$-th level $\Gamma_P^i$ of P is the set of all elements with height $i$, i.e.,
$$\Gamma_P^i = \{a \in P : h(a) = i\}.$$

We remark that each level  $\Gamma_P^i$ is a chain  and the extremal condition of chains can be grasped from the fact that $h(P)=n$ if, and only if, $P$ is a chain or, equivalently, if every level satisfies  $| \Gamma_P^i|=1$. Similarly, for an anti-chain, we have $h(P)=1$ and $| \Gamma_P^1|=n$. If $X\subseteq P$ is a chain or an anti-chain, we say that $X$ is a chain or anti-chain \emph{in} $P$. If $X$ is a chain (anti-chain), we call its cardinality $|X|$ the \emph{length} of the chain (anti-chain).

A broader class of posets, containing both anti-chain and chain posets, is the class of hierarchical posets. We say that $P = ([n], \leqslant_P )$ is \emph{hierarchical} if elements at different levels are always comparable, i.e., $a \leqslant_P b$ for every $a\in\Gamma_P^i$ and $b\in \Gamma_P^j$ where $1 \leq i < j \leq h(P)$. This is an interesting class of posets that will play a central role in the applications to coding theory, as we shall see in Section \ref{sec:applications}.

\medskip
We now show how a poset determines a metric over a finite-dimensional vector space. Due to the primary concern to applications in coding theory, we will restrict ourselves to the case of finite fields, so we let $\mathbb{F}_q$ be the finite field with $q$ elements and $\mathbb{F}_q^n$ be the vector space of all $n$-tuples over $\mathbb{F}_q$. A $k$-dimensional subspace  $\mathcal{C}\subset \mathbb{F}_q^n$ is called an $[n,k]$-\textit{linear code} or, simply a \textit{code} when the parameters are either clear from the context or not relevant.

\medskip

 Given a vector $v=(v_1,\ldots,v_n)\in\mathbb{F}_q^n$, its \emph{support} is the set of all non-null coordinates: $supp(v)=\{i\in [n]:v_i\neq 0\}$.  Given a poset $P$ over $[n]$, the \emph{poset weight} (or $P$\emph{-weight}) of an element $v\in\mathbb{F}_q^n$ is the cardinality of the ideal generated by its support, i.e., $$w_{P}(v)= |\langle supp(v)\rangle_P|.$$
The \emph{poset distance} (or $P$\emph{-distance}) $d_P$ is the distance function over $\mathbb{F}_q^n$ induced by the $P$-weight, in other words, if $u,v\in\mathbb{F}_q^n$, then $$d_P(u,v)=w_P(u-v).$$ The poset distance was introduced by Brualdi et.al. \cite{brualdi1995codes} and it is immediate to prove that $(\mathbb{F}_q^n,d_P)$ is a metric space. The space $(\mathbb{F}_q^n,d_P)$ is called $P$\emph{-space} and the metric $d_P$ is called $P$\emph{-metric}.

A linear \emph{$P$-isometry} is a linear map $T:\mathbb{F}_q^n\rightarrow \mathbb{F}_q^n$ preserving the $P$-weight, and hence the $P$-distance. Two codes $\mathcal{C}_1$ and $\mathcal{C}_2$ over $\mathbb{F}_q^n$ are said to be \emph{$P$-equivalent} if there is a linear isometry $T$ such that $T(\mathcal{C}_1)=\mathcal{C}_2$. 
This is clearly an equivalence relation on the set of all codes and to determine the equivalence classes we need to know what are the linear $P$-isometries of the space.
The group of all linear $P$-isometries over $\mathbb{F}_q^n$ is denoted by $GL_P(\mathbb{F}_q^n)$, and it  is described by the next proposition.

\begin{proposition}[\cite{panek2008groups}]\label{propos1}
	Let $P$ be a poset over $[n]$. Given $T\in GL_P(\mathbb{F}_q^n)$, the map $\phi_T(i)=\mathcal{M}(supp(T(e_i)))$ is a $P$-automorphism. Furthermore, $T\in GL_P(\mathbb{F}_q^n)$ if, and only if,  
	\[
	T(e_i)=\sum_{i\leqslant_P j} x_{ij}e_{\phi_T(i)},
	\]
	where $x_{ij}$ are scalars satisfying $x_{jj}\neq 0$ for every $j\in [n]$.
\end{proposition}

Let $\mathcal{G}_P$ be the subgroup of all linear isometries $T$ such that $\phi_T=id$. Considering the subgroup of linear isometries induced by the automorphisms of $P$ (the permutation part), witch is commonly denoted by $\mathcal{A}ut(P)$, in \cite{panek2008groups}, it was proved that 
\[
GL_P(\mathbb{F}_q^n)=\mathcal{G}_P\rtimes \mathcal{A}ut(P).
\]

\subsection{Partitions and Decompositions}\label{sub:parti}

A \index{partition}\textit{partition} of a subset $J\subseteq\left[  n\right]$ is a family of non-empty subsets $\{J_1,\ldots,J_r\}$ such that  
$$J=\bigcup_{i=1}^{r}J_{i} \text{ and } J_i\cap J_j =\emptyset \text{ for every } i\neq j.$$
We will denote such partition by $\mathcal{J}=(J_{i})_{i=1}^r$. The triple $\mathcal{J}^{\ast}=\left(J;J_0;J_{i}\right)_{i=1}^r$ where 
$J_{0}=\left[  n\right]  \backslash J=\left\{  i\in\left[  n\right]  \ : \ i\notin
J\right\} $ is called \emph{pointed partition}. Since $J$ is the union of the subsets $J_i$, the pointed partition $\mathcal{J}^{\ast}$ is completely determined once the pair $\left(J_0;J_{i}\right)_{i=1}^r$ is given, so we may also write $\mathcal{J}^{\ast}=\left(J_0;J_{i}\right)_{i=1}^r$. Note that $J_{0}=\emptyset$ if, and only if, $J=\left[  n\right]$. We stress that the \emph{pointer} $J_{0}$ has a
special role, since it is the only part we allow to be empty. From now on, we consider only pointed partitions, hence we will omit the symbol $^{\ast}$ and the adjective ``pointed''. 

A partition $\mathcal{J}$  can be
refined in two ways: either by increasing the number of parts or by enlarging
the distinguished part $J_{0}$. Except for the pointer
$J_{0}$, the order of the other parts is irrelevant, for example,
\[
\left(  J_{0};\left\{  1,2\right\}  ,\left\{  3,4,5\right\}  \right)  =\left(
J_{0};\left\{  5,4,3\right\}  ,\left\{  1,2\right\}  \right)  \text{.}%
\]

\begin{definition}
Given a partition $\mathcal{J}=\left(J_{0};J_{i}\right)_{i=1}^{r}$, a \emph{split of} $\mathcal{J}$ is a partition $\mathcal{J}^{\prime
}=(  J_{0};J_{i}^{\prime})_{i=1}^{r+1}$ where $J_{l}=J_{l}^{\prime}\cup J_{r+1}^{\prime}$ for some $l\in[r]$ and $J_{i}=J_{i}^{\prime}$ for each $i\neq l$.
This means that $J_l$ is split into two components and the others are unchanged. 
An \emph{aggregate} of $\mathcal{J}$ is a partition $\mathcal{J}^{\prime}=(  J_{0}^{\prime}%
;J_{i}^{\prime})  _{i=1}^{r}$ where for some $l\in[r]$, $J_{i}^{\prime}=J_{i}$ if $i\not\in\{l,0\}$, $J_{l}=J_{l}^{\prime}\cup J_{l}^{\ast}$ and $J_{0}^{\prime}=J_{0}\cup
J_{l}^{\ast}$ for some $\emptyset\neq J_{l}^{\ast}\varsubsetneq J_{l}$, i.e., some elements of $J_{l}$ were aggregated into the distinguished part $J_{0}$.
\end{definition}

\begin{definition}\label{defdes}
We say that a partition $\mathcal{J}^{\prime}$ is a $1$\emph{-step refinement} of $\mathcal{J}$ if $\mathcal{J}^{\prime}$ is obtained from $\mathcal{J}$
either by a splitting or by aggregating only one element. The partition $\mathcal{J}^{\prime}$ is a \emph{refinement of} $\mathcal{J}$ if $\mathcal{J}^{\prime}$ can be obtained
from $\mathcal{J}$ by a successive number of $1$-step refinements and, if that is the case, we use the
notation $\mathcal{J} \geq \mathcal{J}^{\prime}$.
\end{definition}

\begin{example}\label{exdec}
The partition $([4];\emptyset ; \{1,2,3,4\})$ can be refined in order to get the partition $([4];\{1,3\};\{2\},\{4\})$ by using the following $1$-step refinements:
$$\left(  \emptyset;\left\{  1,2,3,4\right\}  \right)   \geq\left( 
\left\{  3\right\}  ;\left\{  1,2,4\right\}  \right)  \geq\left(
\left\{  1,3\right\}  ;\left\{  2,4\right\}  \right)  \geq\left(
\left\{  1,3\right\}  ;\left\{  2\right\}  ,\left\{  4\right\}  \right).
$$
\end{example}

Set partitions and refinement operations allow us to define decompositions of codes.

\begin{definition}
\label{code_decomposition}We say that $\mathscr{C}=\left(  \mathcal{C};\mathcal{C}_{0};\mathcal{C}_{i}\right)  _{i=1}^{r}$ is a \index{code!decomposition}\emph{decomposition of} an $[n,k,\delta]_q$ code $\mathcal{C}$ if each
$\mathcal{C}_{i}$ is a subspace of $\mathbb{F}_{q}^{n}$ and

\begin{itemize}
\item[(a)] $\mathcal{C}=\oplus_{i=1}^{r}\mathcal{C}_{i}$ with $dim(\mathcal{C}_{i})  >0$ for every $i\in [r]$;
\item[(b)] $\mathcal{C}_{0}=\left\{  \left(  x\,_{1},x_{2},\ldots,x_{n}\right) \ : \ x_{i}=0\text{ if }  i\in supp\left(  \mathcal{C}\right)  \right\}$;
\item[(c)] $\left(supp\left(  \mathcal{C}_{0}\right)  ;supp\left(
\mathcal{C}_{i}\right)  \right)  _{i=1}^{r}$ is a partition over $[n]$.
\end{itemize}
\end{definition}

\begin{definition}
A split, aggregate, $1$-step refinement and a refinement $\mathscr{C}%
^{\prime}=(  \mathcal{C};\mathcal{C}_{0}^{\prime};\mathcal{C}_{i}^{\prime})  _{i=1}^{r\prime}$ of
a decomposition $\mathscr{C}=\left(  \mathcal{C};\mathcal{C}_{0};\mathcal{C}_{i}\right)  _{i=1}^{r}$ are
defined according to
$
(  supp (  \mathcal{C}_{0}^{\prime} )  ;supp(
\mathcal{C}_{i}^{\prime})  )  _{i=1}^{r\prime}%
$
being a split, aggregate, $1$-step refinement or a refinement of
$\left(  supp\left( \mathcal{C}_{0}\right)  ;supp\left(
\mathcal{C}_{i}\right)  \right)  _{i=1}^{r}$ respectively. We will also use the notation $\mathscr{C}\geq \mathscr{C}^\prime$ to denote a refinement.
\end{definition}

\begin{example}
	Let $P$ be a poset over $[4]$. Consider the $[4,2,\delta_P]_2$ code $\mathcal{C}$ given by
	\[
	 	\mathcal{C}=\{0000,1100,0010,1110\}.
	  \] 
	  Then,
	  \[
	  	(\mathcal{C}; \langle e_4 \rangle; \mathcal{C})\geq (\mathcal{C}; \langle e_4 \rangle; \mathcal{C}_i)_{i=1}^2  
	  \]
	  where
	  \[
	  	\mathcal{C}_1=\{0000,1100\} \text{ and } \mathcal{C}_2=\{0000,0010\}.
 	  \]
\end{example}

\begin{definition}
	A decomposition $\mathscr{C}=(\mathcal{C};\mathcal{C}_0;\mathcal{C}_i)_{i=1}^r$ is said to be \index{maximal decomposition}\textit{maximal} if it does not admit a refinement.
\end{definition}

Until now, only the algebraic aspects of the codes were used, the metric $d_{P}$ has not played any role in the decomposition of a code. We introduce now a decomposition that depends of the metric and consequently of the poset $P$.  

\begin{definition}
\label{P_decomposition} A \index{code!$P$-decomposition}$P$-\emph{decomposition of} $\mathcal{C}$ is a decomposition 
$\mathscr{C}=(  \mathcal{C}^{\prime};\mathcal{C}_{0}^{\prime};\mathcal{C}_{i}^{\prime
})  _{i=1}^{r}$ of
$\mathcal{C}^{\prime}$ (as in Definition \ref{code_decomposition}) where $\mathcal{C}^{\prime}\sim_{P}\mathcal{C}$. Each $\mathcal{C}_{i}^{\prime}$ is called a \emph{component }of the decomposition. A 
\emph{ trivial }$P$\emph{-decomposition} of $\mathcal{C}$ is either the decomposition
$\left(  \mathcal{C};\mathcal{C}_{0};\mathcal{C}\right)  $ or any $P$-decomposition with a unique factor
$(  \mathcal{C}^{\prime};\mathcal{C}_{0}^{\prime};\mathcal{C}^{\prime})  $ where $\vert
supp(  \mathcal{C}^{\prime})  \vert =\left\vert supp%
\left(  \mathcal{C}\right)  \right\vert $ and $\vert supp(
\mathcal{C}_{0}^{\prime})  \vert =\left\vert supp\left(
\mathcal{C}_{0}\right)  \right\vert $.
\end{definition}

\begin{definition}
A code $\mathcal{C}$ is said to be $P$\emph{-irreducible} if it does not admit a non-trivial $P$-decomposition.
\end{definition}

		

\begin{example}\label{exemplo3}
	Let $P$ be the hierarchical poset with order relations $1,2,3\leqslant 4$. The codes 
\[
	\mathcal{C}=\{0000,1001\} \ \text{ and }\ \mathcal{C}^\prime= \{0000,0111\}
\]
are equivalent since
\[
	T(e_1)=e_1, \ T(e_2)=e_2, \ T(e_3)=e_3, \ T(e_4)=e_1+e_2+e_3+e_4
\]
is an isometry satisfying $T(\mathcal{C})=\mathcal{C}^\prime$. Then, $\mathscr{C}^\prime=(\mathcal{C}^\prime; \langle e_1 \rangle; \mathcal{C}^\prime)$ is a $P$-decomposition of $\mathcal{C}$ while $\mathscr{C}=(\mathcal{C};\langle e_2, e_3 \rangle; \mathcal{C})$ is a decomposition of $\mathcal{C}$. 
\end{example}

We note that if we consider the partitions $([4];\{2,3\};\{1,4\})$ and $([4];\{1\};\{2,3,4\})$ induced by the $P$-decompositions $\mathscr{C}$ and $\mathscr{C}^\prime$ constructed in the previous example, these partitions can not be obtained from each other by a sequence of aggregations and splittings. Hence, we must use the order automorphism induced by the isometries in order to compare them. We now do it explicitly.

Let $\mathscr{C}=(
\mathcal{C}^{\prime\prime};\mathcal{C}_{0}^{\prime\prime};\mathcal{C}_{i}^{\prime\prime})  _{i=1}^{r}$ and
$\mathscr{C}^{\prime}=(  \mathcal{C}^{\prime};\mathcal{C}_{0}^{\prime
};\mathcal{C}_{i}^{\prime})  _{i=1}^{r^{\prime}}$ be two $P$-decompositions of a code $\mathcal{C}$. Associated to those
$P$-decompositions there are two  partitions of $\left[  n\right]  $,
namely: $(  supp(  \mathcal{C}_{0}^{\prime\prime})  ;supp%
(  \mathcal{C}_{i}^{\prime\prime}))  _{i=1}^{r}$ and $(supp(  \mathcal{C}_{0}^{\prime})  ;supp(\mathcal{C}_{i}^{\prime}))  _{i=1}^{r^{\prime}}$.
By the definition of a $P$-decomposition, there are isometries $T_1,T_2\in GL_{P}(\mathbb{F}_q^n)$ such that $T_1\left(  \mathcal{C}\right)  =
\mathcal{C}^{\prime\prime}$ and $T_2\left(  \mathcal{C}\right)  =\mathcal{C}^{\prime}$. Denote $T=T_2\circ\left(  T_1\right)  ^{-1}$. Then $T$ is a linear isometry and $T(\mathcal{C}^{\prime\prime})=\mathcal{C}^{\prime}$. Considering the automorphism $\phi_T$ of Proposition \ref{propos1} (the order automorphism induced by $T$), 
$\phi_{T}$ induces a map on the  partition of $\left[  n\right]  $ determined by the $P$-decomposition $\mathscr{C}$, namely,
$$\phi_{T}[  (  supp(  \mathcal{C}_{0}^{\prime\prime})
;supp (  \mathcal{C}_{i}^{\prime\prime}))  _{i=1}^{r}]
=  \left(  \phi_{T}\left(  supp\left(  \mathcal{C}_{0}\right)
\right)  ;\phi_{T}\left(  supp\left(  \mathcal{C}_{i}\right)
\right)  \right)  _{i=1}^{r}.$$

Considering the map $T$ defined in Example \ref{exemplo3}, we may define $T_1=\text{identity}$ and $T_2=T$. Then,
$$\phi_{T} \left[([4];\{2,3\};\{1,4\})\right]=([4];\phi_{T}(\{2,3\});\phi_{T}(\{1,4\}))=([4];\{2,3\};\{1,4\})$$
is a partition which also can not be obtained by $([4];\{1\};\{2,3,4\})$ using an aggregation. Considering the isometry 
\[
	S(e_1)=e_3, \ S(e_2)=e_2, \ S(e_3)=e_1, \ S(e_4)=e_2+e_4
\]
we still have that $S(\mathcal{C}^{\prime\prime})=\mathcal{C}^\prime$ but now
$$\phi_{S} \left[([4];\{2,3\};\{1,4\})\right]=([4];\phi_{S}(\{2,3\});\phi_{S}(\{1,4\}))=([4];\{1,3\};\{2,4\})$$
is an aggregation of $([4];\{1\};\{2,3,4\})$. Hence, it may happen that $\mathscr{C}^{\prime}$ and $\mathscr{C}^{\prime\prime}$ may not be comparable for some $U\in GL_P(\mathbb{F}_q^n)$ but comparable for some $T\in GL_P(\mathbb{F}_q^n)$. Therefore, we can define the analogous of the operations on decompositions by  demanding the existence of such an isometry. 

\begin{definition}
Let $\mathscr{C}^{\prime}=(  \mathcal{C}^{\prime};\mathcal{C}_{0}^{\prime}%
;\mathcal{C}_{i}^{\prime})  _{i=1}^{r\prime}$ and $\mathscr{C}^{\prime\prime
}=(  \mathcal{C}^{\prime\prime};\mathcal{C}_{0}^{\prime\prime};\mathcal{C}_{i}^{\prime\prime})
_{i=1}^{r\prime\prime}$ be two $P$-decompositions of $\mathcal{C}$. We say that
$\mathscr{C}^{\prime}$ is \emph{a }$P$\emph{-refinement} of $\mathscr{C}^{\prime\prime}$ (and write $\mathscr{C} \geq \mathscr{C}^\prime$) if there is  $T\in GL_P(\mathbb{F}_q^n)$ such that $\phi_{T}[
(  supp(  \mathcal{C}_{0}^{\prime})  ;supp(
\mathcal{C}_{i}^{\prime}))  _{i=1}^{r\prime}]  $ is a refinement
of the partition $ (  supp(
\mathcal{C}_{0}^{\prime\prime})  ;supp(  \mathcal{C}_{i}^{\prime\prime
})  )  _{i=1}^{r^{\prime\prime}}$. 
\end{definition}

Since, for every $T\in GL_P(\mathbb{F}_q^n)$, the induced map $\phi_T$ is an automorphism, if $\mathscr{C}$ and $\mathscr{C}^\prime$ are comparable by $S$ and $T$ respectively, the number of aggregations and splittings used to obtain $\mathscr{C}^{\prime}$ from $\mathscr{C}^{\prime\prime}$ coincides when $S$ or $T$ are used to compare them. Hence, splittings and aggregations are well determined once two $P$-decompositions are given.


	

\begin{definition}\label{com}
A $P$-decomposition $\mathscr{C}=\left(  \mathcal{C};\mathcal{C}_0;\mathcal{C}_{i}\right)_{i=1}^r$ is said to be \index{$P$-decomposition!maximal}\emph{maximal} if it can not be refined or, equivalently, if each $\mathcal{C}_{i}$ is $P$-irreducible for every $i\in[r]$.
\end{definition}

Let us consider the Hamming metric $d_{H}$ over $\mathbb{F}_2^n$. It is well-known that the group of linear isometries of this metric space is isomorphic to the permutation group $\mathcal{S}_{n}$. Given a code $\mathcal{C}$, let
$\mathscr{C}=(\mathcal{C};\mathcal{C}_0;\mathcal{C}_i)_{i=1}^r$ be a maximal decomposition of $\mathcal{C}$. If a code $\mathcal{C}^{\prime}$
is $d_H$-equivalent to $\mathcal{C}$, then there is $T\in GL_{d_H}(\mathbb{F}_q^n)\sim \mathcal{S}_n$ such that $T\left(
\mathcal{C}\right)  =\mathcal{C}^{\prime}$. Note that the decomposition $\mathscr{C}^\prime=(\mathcal{C}^\prime;T(\mathcal{C}_0);T(\mathcal{C}_i))_{i=1}^r$ is a maximal decomposition of
$\mathcal{C}^{\prime}$, otherwise, $\mathscr{C}$ would not be a maximal decomposition of $\mathcal{C}$. Therefore, when considering the Hamming metric, a maximal decomposition is also a maximal $H$-decomposition ($H$ is the anti-chain poset over $[n]$). This is not true in general, as we can see in Example \ref{exemplo3}.


\begin{definition}
Let $\mathscr{C}=\left(
\mathcal{C}^{\prime};\mathcal{C}_{0};\mathcal{C}_{i}\right)  _{i=1}^{r}$ be a
$P$-decomposition of an $\left[  n,k,\delta\right]  _{q}$ code $\mathcal{C}$. The
\index{$P$-decomposition!profile}\emph{profile} of
$\mathscr{C}$ is the array 
\[ \mathrm{profile}\left(  \mathscr{C}\right) :=\left[  \left(  n_{0}%
,k_{0}\right)  ,\left(  n_{1},k_{1}\right)
,\ldots,\left(  n_{r},k_{r}\right)  \right],  
\]
where
\[
n_{i}=\left\vert supp\left(  \mathcal{C}_{i}\right)
\right\vert \text{ and }k_{i}=dim\left(  \mathcal{C}_{i}\right)
\text{.}%
\]
\end{definition}

Let $\beta=\left\{  e_{1},\ldots,e_{n}\right\}  $ be the canonical basis of $\mathbb{F}_q^n$. Given $I\subset\left[
n\right]$, the $I$-\emph{coordinate subspace} $V_I$ is defined by    
\[
V_I=span \left\{e_i \ : \ i\in I\right\} =\left\{  \sum_{i\in I}x_{i}e_{i} \ : \ x_i\in \mathbb{F}_q\right\} .
\]
Given a decomposition $\mathscr{C}=(\mathcal{C};\mathcal{C}_0;\mathcal{C}_i)_{i=1}^r$ of a linear code $\mathcal{C}$, we say that 
\begin{center}
$
V_{i}:= V_{supp(\mathcal{C}_i)}=span \left\{e_i \ : \ i\in supp\left(
\mathcal{C}_{i}\right)\right\}  $
\end{center}
is the \emph{support-space of (the component) }$\mathcal{C}_{i}$. We consider
$\left[  n\right]  _{\mathcal{C}}=supp\left(  \mathcal{C}\right)  $ and $\left[
n\right]  ^{\mathcal{C}}=\left[  n\right]  \backslash\left[  n\right]  _{\mathcal{C}}$. In the case where $\left[  n\right]  ^{\mathcal{C}}\neq\emptyset$, we  write 
$V_{0}=V_{[n]^{\mathcal{C}}}$ and denote $\mathcal{C}_{0}=V_{0}$. We
say that the decomposition $(\mathcal{C};\mathcal{C}_0;\mathcal{C}_i)_{i=1}^r$ is supported by the
\emph{environment decomposition }$(V_0;V_i)_{i=1}^r$. In the case where $\left[
n\right]  ^{\mathcal{C}}=\emptyset$, we have $V_{0}=\mathcal{C}_{0}=\left\{  \mathbf{0}\right\}  $.

Given a $P$-decomposition $\mathscr{C}=(  \mathcal{C}^\prime;\mathcal{C}_{0}^{\prime};\mathcal{C}_{i}^{\prime})
_{i=1}^{r}$  of $\mathcal{C}$, the environment decomposition determine a partition of $[n]$, i.e., $\left[  n\right]  =\cup_{i=0}^{r}supp\left(  V_{i}\right)  $. Also, if $\mathcal{C}_{i}^\prime=\left\{  \mathbf{0}\right\}$, then $ i=0$. If each $\mathcal{C}^\prime_i$ is $P$-irreducible, denoting $n_{i}=dim (V_{i})=|supp(\mathcal{C}_i)|$
and $k_{i}=dim (\mathcal{C}_{i}^{\prime})$, then 
\begin{equation*}
\sum_{i=0}^{r}n_{i}   =n=dim (\mathbb{F}_q^n )\ \text{ and } \ 
\sum_{i=1}^{r}k_{i}=k=dim (\mathcal{C}).
\end{equation*}

The following theorem states that the profile of a maximal $P$-decomposition
$\mathscr{C}$ of a code $\mathcal{C}$ depends (essentially) exclusively on
$\mathcal{C}$, not on $\mathscr{C}$.

\begin{theorem}\label{teo:profile}
\label{teo1}Let $\mathcal{C}$ be an $\left[  n,k,\delta\right]  _{q}$ code and let $P$
be a poset over $\left[  n\right]  $. If $\mathscr{C}^{\prime}$ and
$\mathscr{C}^{\prime\prime}$ are two maximal $P$-decompositions of $\mathcal{C}$ with%
$$
\mathrm{profile}(  \mathscr{C}^{\prime})    =\left[  (
n_{0}^{\prime},k\,_{0}^{\prime})  ,(  n_{1}^{\prime},k_{1}^{\prime
})  ,\ldots,(  n_{r}^{\prime},k_{r}^{\prime})  \right] 
$$
and
$$
\mathrm{profile}(  \mathscr{C}^{\prime\prime})   =\left[
(  n_{0}^{\prime\prime},k_{0}^{\prime\prime})  ,(
n_{1}^{\prime\prime},k_{1}^{\prime\prime})  ,\ldots,(  n_{s}%
^{\prime\prime},k_{s}^{\prime\prime})  \right]  \text{.}%
$$
Then, $r=s$ and, up to a permutation, $\mathrm{profile}\left(  \mathscr{C}%
^{\prime}\right)  =\mathrm{profile}\left(  \mathscr{C}^{\prime\prime}\right)
$, i.e., there is $\sigma\in \mathcal{S}_{r}$ such that $(  n_{i}^{\prime}%
,k_{i}^{\prime})  =(  n_{\sigma(  i)  }^{\prime\prime
},k_{\sigma(  i)  }^{\prime\prime})  $ and $(
n_{0}^{\prime},k_{0}^{\prime})  =(  n_{0}^{\prime\prime}%
,k_{0}^{\prime\prime})  $.
\end{theorem}

\begin{IEEEproof}
	Let $\mathscr{C}^{\prime}=(  \mathcal{C}^{\prime};\mathcal{C}_{0}^{\prime
	};\mathcal{C}_{i}^{\prime})  _{i=1}^{r}$ and $\mathscr{C}^{\prime\prime}=(
	\mathcal{C}^{\prime\prime};\mathcal{C}_{0}^{\prime\prime};\mathcal{C}_{i}^{\prime\prime})  _{i=1}^{s}$
	be two maximal $P$-decompositions of $\mathcal{C}$. Suppose, without loss of generality, $r<s$. If $T\in GL_{P}(\mathbb{F}_q^n)$
	is an isometry satisfying $T\left(  \mathcal{C}^{\prime}\right)  =\mathcal{C}^{\prime\prime}$, there is a
	component $\mathcal{C}_{i}^{\prime}$ of $\mathscr{C}^{\prime}$ such that $T(  \mathcal{C}_{i}^{\prime
	})  $ is not contained in any component $\mathcal{C}_{j}^{\prime\prime}$ of
	$\mathscr{C}^{\prime\prime}$, otherwise $r\geq s$. Hence, there are components $\mathcal{C}_{i_{0}}^{\prime}$ of $\mathscr{C}^\prime$ and $\mathcal{C}_{j_{0}}%
	^{\prime\prime},\mathcal{C}_{j_{1}}^{\prime\prime},\ldots,\mathcal{C}_{j_{t}}^{\prime\prime}$ of $\mathscr{C}^{\prime\prime}$ such that 
	\[
	T(  \mathcal{C}_{i_{0}}^{\prime})  \subset \mathcal{C}_{j_{0}}^{\prime\prime}\oplus
	\mathcal{C}_{j_{1}}^{\prime\prime}\oplus\cdots\oplus \mathcal{C}_{j_{t}}^{\prime\prime}%
	\]
	and $T(  \mathcal{C}_{i_{0}}^{\prime}) \cap \mathcal{C}_{j_{l}}^{\prime\prime} \neq \emptyset$
	for any $l\in\{1,\ldots,t\}$. Therefore, 
	\[
	T(  \mathcal{C}_{i_{0}}^{\prime})  =\bigoplus_{m=0}^t T(  \mathcal{C}_{i_{0}}^{\prime
	})  \cap \mathcal{C}_{j_{m}}^{\prime\prime}
	\]
	is a non-trivial $P$-decomposition for $\mathcal{C}_{i_0}^\prime$, contradicting the fact that each component of a maximal $P$-decomposition is $P$-irreducible. It follows that $r=s$. Moreover, 
	for every $i\in\{1,\ldots,r\}$ there is $j_{i}$ such that $T(
	\mathcal{C}_{i}^{\prime})  \subseteq \mathcal{C}_{j_{i}}^{\prime\prime}$. Hence, 
	$n_{i}^{\prime}\leq$ $n_{j_{i}}^{\prime\prime}$ and $k_{i}^{\prime}\leq
	k_{j_{i}}^{\prime\prime}$. Applying the same reasoning to $T^{-1}\in
	GL_{P}(\mathbb{F}_q^n)$, we get that $n_{i}^{\prime\prime}\leq$ $n_{j_{i}%
	}^{\prime}$ and $k_{i}^{\prime\prime}\leq k_{j_{i}}^{\prime}$, hence
	$n_{i}^{\prime}=$ $n_{j_{i}}^{\prime\prime}$ and $k_{i}^{\prime}=k_{j_{i}%
	}^{\prime\prime}$, so that, up to a permutation, $\mathrm{profile}(  \mathscr{C}^{\prime
})  =\mathrm{profile}(  \mathscr{C}^{\prime\prime})  $.
\end{IEEEproof}
The next Corollary follows straight from Theorem \ref{teo1}.
\begin{corollary}\label{cor:permut}
	Let $\mathscr{C}^{\prime}=(  \mathcal{C}^{\prime};\mathcal{C}_{0}^{\prime};\mathcal{C}_{i}^{\prime
	})  _{i=1}^{r}$ and $\mathscr{C}^{\prime\prime}=(  \mathcal{C}^{\prime\prime
	};\mathcal{C}_{0}^{\prime\prime};\mathcal{C}_{i}^{\prime\prime})_{i=1}^{r}$ be two maximal
	$P$-decompositions of $\mathcal{C}$ and let $T\in GL_{P}(\mathbb{F}_q^n)$ be a linear
	isometry such that $T\left(  \mathcal{C}^{\prime}\right)  =\mathcal{C}^{\prime\prime}$. Then,
	there is a permutation $\sigma\in \mathcal{S}_{r}$ such that $T(  \mathcal{C}_i^{\prime})  =\mathcal{C}_{\sigma\left(  i\right)  }^{\prime\prime}$.
\end{corollary}

To express the number of operations (splitting and aggregations) performed in a decomposition from a trivial decomposition, we first need to show that every non-trivial $P$-decomposition is a refinement of a trivial $P$-decomposition.

\begin{lemma}\label{le11}
	If $T\in GL_P(\mathbb{F}_q^n)$ then the ideals $\langle supp(\mathcal{C})\rangle$ and $\langle supp(T(\mathcal{C}))\rangle$ are isomorphic.
\end{lemma}
\begin{IEEEproof}
	Given $T\in GL_P(\mathbb{F}_q^n)$, follows immediate from the definition of $\phi_T$ (Proposition \ref{propos1}) that 
\[
	\phi_T(\mathcal{M}(supp(\mathcal{C})))=\mathcal{M}(supp(T(\mathcal{C}))).
\]
Hence, 
\[
	\langle supp(T(\mathcal{C}))\rangle=\langle \mathcal{M}(supp(T(\mathcal{C})))\rangle=\langle \phi_T(\mathcal{M}(supp(\mathcal{C}))) \rangle.
\]
On the other hand, since $\phi_T$ is an isomorphism,
\[
	\langle supp(\mathcal{C})\rangle=\langle \mathcal{M} (supp(\mathcal{C}))\rangle \sim \phi_T (\langle \mathcal{M} (supp(\mathcal{C}))\rangle ) = \langle \phi_T (\mathcal{M}(supp(\mathcal{C})))\rangle.
\]
Therefore,
\[
 	\langle supp(\mathcal{C})\rangle \sim \langle supp(T(\mathcal{C}))\rangle.
 \] 
\end{IEEEproof}

\begin{proposition}
	If $\mathscr{C}=(\mathcal{C}^\prime;\mathcal{C}_0^\prime;\mathcal{C}_i^\prime)_{i=1}^r$ is a non-trivial $P$-decomposition of $\mathcal{C}$ with $|supp(\mathcal{C}_0^\prime)|>|supp(\mathcal{C}_0)|$, 
then $\mathscr{C}$ is a refinement of a trivial $P$-decomposition of $\mathcal{C}$.
\end{proposition}

\begin{IEEEproof}
	  It is straightforward that $\mathscr{C}$ is a refinement of the one-component $P$-decomposition $(\mathcal{C}^\prime;\mathcal{C}_0^\prime;\mathcal{C}^\prime)$. To conclude, we need to prove that $(\mathcal{C}^\prime;\mathcal{C}_0^\prime;\mathcal{C}^\prime)$ is obtained by aggregations from a trivial $P$-decomposition. Since $|supp(\mathcal{C}^\prime)|<|supp(\mathcal{C})|$ and $|\mathcal{M}(supp(\mathcal{C}^\prime))|=|\mathcal{M}(supp(\mathcal{C}))|$, there exist $i_0\in \langle supp(\mathcal{C}^\prime)\rangle\setminus supp(\mathcal{C}^\prime)$, otherwise $\langle supp(\mathcal{C})\rangle\not\sim \langle supp(\mathcal{C}^\prime)\rangle$ contradicting Lemma \ref{le11}. Take $j_0\in \mathcal{M}(supp(\mathcal{C}^\prime))$ such that $j_0>_P i_0$ and define $S\in GL_P(\mathbb{F}_q^n)$ by
\[
	S(e_i)=e_i \ \forall \ i\neq j_0 \ \ \ \text{ and} \ \ \ S(e_{j_0})=e_{j_0}+e_{i_0}.
\]
Therefore, $\phi_S=id$. Denoting $S(\mathcal{C}^\prime)=\mathcal{C}^{\prime\prime}$, since $supp(\mathcal{C}^\prime)=supp(\mathcal{C}^{\prime\prime})\cup \{i_0\}$ we get that $(\mathcal{C}^\prime;\mathcal{C}_0^\prime;\mathcal{C}^\prime)$ is a 1-step refinement of $(\mathcal{C}^{\prime\prime};\mathcal{C}_0^{\prime\prime};\mathcal{C}^{\prime\prime})$. Proceeding in this form (performing aggregations), at some point we get a trivial $P$-decomposition having $\mathscr{C}$ as its refinement.
\end{IEEEproof}

\begin{definition}
	Given a $P$-decomposition $\mathscr{C}=(\mathcal{C}^\prime; \mathcal{C}_0^\prime; \mathcal{C}_i^\prime)_{i=1}^r$ of $\mathcal{C}$, its \textit{degree} according to $\mathcal{C}$ is defined by 
	\[
		\mathcal{D}_P(\mathscr{C},\mathcal{C})=(r-1)+ |supp(\mathcal{C}_0^\prime)|-|supp(\mathcal{C}_0)|.
	\]	
	A $P$-decomposition with maximum degree is called \textit{primary $P$-decomposition}. 
\end{definition}

The degree of a decomposition is basically the number of $1$-step refinements necessary to obtain the given decomposition from a trivial decomposition, clearly, if $\mathscr{C}$ is a trivial decomposition, $\mathcal{D}_P(\mathscr{C},\mathcal{C})=0$ and if $\mathscr{C}^\prime$ is a $P$-refinement of $\mathscr{C}$, then $\mathcal{D}_P(\mathscr{C}^\prime,\mathcal{C})>\mathcal{D}_P(\mathscr{C},\mathcal{C})$. Furthermore, the degree of a decomposition is completely determined by its profile. Hence, by Theorem \ref{teo1}, maximal $P$-decompositions have the same degree. Thus, the maximum degree of a decomposition of a code $\mathcal{C}$ will be denoted by $\mathcal{D}_P(\mathcal{C})$ instead of $\mathcal{D}_P(\mathscr{C},\mathcal{C})$.

The following proposition follows straight from the previous comments.



\begin{proposition}\label{prop:maximal_primary}
	A $P$-decomposition of a code $\mathcal{C}$ is maximal if, and only if, it is a primary $P$-decomposition.
\end{proposition}

We recall that the group of linear $P$-isometries is the semi-direct product $GL_P(\mathbb{F}_q^n)=\mathcal{G}_P\rtimes \mathcal{A}ut(P)$. It is worth remarking that the permutation part $\mathcal{A}ut(P)$ is irrelevant regarding maximality of $P$-decompositions:


\begin{proposition}
	\label{lema1}Let $\mathscr{C}=\left(  \mathcal{C}^{\prime}%
	;\mathcal{C}_{0};\mathcal{C}_{i}\right)_{i=1}^r$ be a maximal
	$P$-decomposition of $\mathcal{C}$. Let $\phi\in Aut\left(  P\right)  $ and $T_{\phi}\in
	\mathcal{A}ut(P)$ be the isometry induced by $\phi$. Then,
	\[
	\mathscr{C}^{\prime}  =\left(  T_{\phi}\left(
	\mathcal{C}^{\prime}\right)  ;T_{\phi}\left(  \mathcal{C}_{0}\right)  ;T_{\phi}\left(  \mathcal{C}_{i}\right)  \right)_{i=1}^r
	\]
	is also a maximal $P$-decomposition of $\mathcal{C}$.
\end{proposition}

\begin{IEEEproof}
	Because $\phi$ is a permutation, for every $i\in\{0,1,\ldots,r\}$, 
	$$j\in supp(\mathcal{C}_i) \iff \phi(j) \in supp(T_\phi (\mathcal{C}_i)).$$
	Hence, $\mathscr{C}^{\prime}$ is a $P$-decomposition of $\mathcal{C}$. Since its profile coincides with the profile of $\mathscr{C}$, $\mathscr{C}^\prime$ is also a maximal $P$-decomposition of $\mathcal{C}$.
\end{IEEEproof}


\subsection{Construction of Maximal P-Decompositions}

In order to find maximal $P$-decompositions of a code $\mathcal{C}$, we need to obtain the maximal decompositions of its equivalent codes. 
In this section, we will describe the construction of a generator matrix which provides a code equivalent to $\mathcal{C}$ and determines a maximal $P$-decomposition of $\mathcal{C}$. 

Let $P$ be a poset over $[n]$ and $G=(g_{ij})$ be a $k\times n$ generator matrix of an $[n,k,\delta]_{q}$ code $\mathcal{C}$. We lose no generality by assuming that $G$ is in a reduced row echelon form, obtained by elementary operations on rows.
In order to obtain a maximal $P$-decomposition, we need to use a slight different definition for the classical reduced row echelon form. For each $i\in [k]$, let
$$j\left(  i\right)  :=\max\left\{  j \ : \ g_{ij}\neq0\right\} $$
be the right-most non-zero column of the $i$-th row of $G$. Performing elementary row operations on $G$, we may assume that
\begin{equation}\label{eqelemen}
j(1)>j(2)>\cdots > j(k) \ \text{ and } \ g_{ij(l)}=0 \ \text{ if } \ i\neq l.	
\end{equation}
We say that $G$ is in \textit{inverse reduced row echelon form} if the entries of $G$ satisfy Property (\ref{eqelemen}). From now on, we assumed that generator matrices have this form. In order to make more clear the difference of the proposed reduced row echelon form and the classical one, we present the next example.


\begin{example}\label{exxx01}
	Let 
	$$G=\left[
	\begin{array}{ccccc}
	1 & 0 & 1 & 1 & 0\\
	1 & 1 & 0 & 1 & 1\\
	0 & 1 & 0 & 1 & 1 
	\end{array}\right]
	$$
	be a generator matrix of a $[5,3]_2$ code $\mathcal{C}$ (the metric here is irrelevant). Then,
	$$G_1=\left[
	\begin{array}{ccccc}
	1 & 0 & 0 & 0 & 0\\
	0 & 1 & 0 & 1 & 1\\
	0 & 0 & 1 & 1 & 0 
	\end{array}\right] \text{ and }
	G_{2}=\left[
	\begin{array}{ccccc}
	0 & 1 & 1 & 0 & 1\\
	0 & 0 & 1 & 1 & 0\\
	1 & 0 & 0 & 0 & 0 
	\end{array}\right]
	$$
	are also generator matrices of $\mathcal{C}$. Note that $G_1$ is in the classical reduced row echelon form while $G_2$ is in the proposed one. 
	
\end{example} 



A generator matrix $G$ of $\mathcal{C}$ determines a unique decomposition $\mathscr{C}$ in the following sense:

\vspace{0.4cm}
\noindent\textbf{Construction of $\mathscr{C}$:}
\vspace{0.4cm}
\hrule
\vspace{0.3cm}
Suppose $\beta_1=\{v_1,\ldots,v_k\}$ is the set of all rows of $G$ and $I\subset [n]$ is the index set of the null columns of $G$. Then, define 
\[
\mathcal{C}_0=\left\{v\in \mathbb{F}_q^n \ : \ v=\sum_{i\in I} x_i e_i\right\}.
\]
Take $w_1\in \beta_1$ and let $\gamma_1=\{v_{i_1},\ldots,v_{i_r}\}\subset \beta_1$ be the set of all rows of $G$ such that if $v\in\gamma_1$, then $supp(w_1)\cap supp(v)\neq \emptyset$. Denote 
$$\mathcal{C}_1=\left\{c\in \mathcal{C} \ : \ c=\sum_{j=1}^r x_i v_{i_j}\right\}.$$
Take $\beta_2=\beta_1\setminus \gamma_1$, if it is empty, the decomposition $(\mathcal{C};\mathcal{C}_0;\mathcal{C}_1)$ is the one determined by $G$. If $\beta_2\neq \emptyset$, take $w_2\in \beta_2$ and $\gamma_2=\{v_{i_1},\ldots,v_{i_s}\}\subset \beta_2$ where $v\in \gamma_2$ if, and only if, $supp(w_2)\cap supp(v)\neq \emptyset$. Define
\[
\mathcal{C}_2=\left\{c\in \mathcal{C} \ : \ c=\sum_{j=1}^s x_i v_{i_j}\right\}.	
\] 
Proceeding in this way, $\beta_{r+1}=\emptyset$ and $\beta_r\neq \emptyset$ for some $r$. In this case, $\mathscr{C}=(\mathcal{C};\mathcal{C}_0;\mathcal{C}_i)_{i=1}^r$ is the decomposition determined by $G$. It is clear that the choice of $w_i$ does not change the degree of the decomposition, therefore, the decomposition constructed is unique, up to a permutation of its components. As we shall see in the next proposition, this decomposition is maximal.
\vspace{0.3cm}
\hrule
\vspace{0.6cm}

\begin{example}
	Consider the generator matrices of Example \ref{exxx01}. The decomposition determined by $G$ is trivial since every two rows of $G$ have intersection in their supports. On the other hand, the decomposition obtained by the matrix $G_2$ is not trivial, indeed, the third row has disjoint support from the first and second rows. Therefore, if 
	\[
	\mathcal{C}_1^\prime=\{00000,01011,00110,01101\}, \ \ 
	\mathcal{C}_2^\prime=\{00000,10000\},
	\]
	and
	\[
	\mathcal{C}_1=\{00000,01101,00110,01011\}, \ \ 
	\mathcal{C}_2=\mathcal{C}_2^\prime,
	\]
	then $(\mathcal{C};\emptyset;\mathcal{C}_i^\prime)_{i=1}^2$ and $(\mathcal{C};\emptyset;\mathcal{C}_i)_{i=1}^2$ are the decomposition obtained from $G_1$ and $G_2$ respectively. 
\end{example}

A generator matrix $G$ is said to be in \index{generalized reduced row echelon form}\textit{generalized reduced row echelon form} if there is a permutation $\sigma \in \mathcal{S}_k$ such that
\[
j(\sigma(1))>\ldots>j(\sigma(k)) \text{ and } g_{ij(l)}=0 \text{ if } i\neq l,
\]
i.e., if a matrix in inverse reduced row echelon form may be obtained by permuting the rows of $G$.

\begin{proposition}\label{maxrow}
	Let $\mathcal{C}$ be an $[n,k,\delta]_q$ code. If $G$ is a generator matrix of $\mathcal{C}$ in a generalized reduced row echelon form, then $G$ determines a maximal decomposition for $\mathcal{C}$.
\end{proposition}

\begin{IEEEproof}
	Let $G$ be a generator matrix of $\mathcal{C}$ in a generalized reduced row echelon form. Without loss of generality, we will assume that the decomposition determined by $G$ has only one component, i.e., $\mathscr{C}=(\mathcal{C}; \mathcal{C}_0; \mathcal{C})$. By the construction of $\mathscr{C}$ it is straightforward that $\mathscr{C}$ can not be refined by aggregations. 
	Suppose $\mathscr{C}^\prime=(\mathcal{C};\mathcal{C}_0;\mathcal{C}_i)_{i=1}^2$ is a splitting of $\mathscr{C}$. Thus, $\mathcal{C}=\mathcal{C}_1 \oplus \mathcal{C}_2$ and $supp(\mathcal{C}_1)\cap supp(\mathcal{C}_2)=\emptyset $. Let $\{w_1,\ldots,w_{k_1}\}$ and $\{w_{k_1+1},\ldots,w_k\}$ be a basis of $\mathcal{C}_1$ and $\mathcal{C}_2$ respectively. The $k\times n$ matrix $G_1$ having $\{w_1,\ldots,w_k\}$ as its rows is the generator matrix of $\mathcal{C}$ whose decomposition induced is $\mathscr{C}^\prime$. Then, 
	each row of $G_1$ is a linear combination of the rows of $G$. Let $\beta_1$ be the minimum set of rows of $G$ generating $\{w_1,\ldots,w_{k_1}\}$, i.e., for every $i\in\{1,\ldots,k_1\}$,	
	\[
	w_i=\sum_{v\in\beta_1} x_v^i v
	\] 
	where $x_v^i\in\mathbb{F}_q$. Similarly, let $\beta_2$ be the minimum set of rows of $G$ generating $\{w_{k_1+1},\ldots,w_{k}\}$. 
	Since $G$ and $G_1$ generates the same code, $\beta_1\cap \beta_2=\emptyset$. 
	Since the decomposition induced by $G$ is trivial, $supp(\beta_1)\cap supp(\beta_2)\neq \emptyset$. Take $i_0\in supp(\beta_1)\cap supp(\beta_2)$, since $span\{\beta_1\}=\mathcal{C}_1$ and $span\{\beta_2\}=\mathcal{C}_2$, it follows that  $i_0\in supp(\mathcal{C}_1)\cap supp(\mathcal{C}_2)$, a contradiction.
\end{IEEEproof}

The previous proposition ensures that each maximal decomposition is obtained by taking a generator matrix in a generalized reduced row echelon form. 
Till now, we have not used the group of linear isometries to construct maximal $P$-decompositions. In the following, we will consider the subgroup of isometries $\mathcal{G}_P$ (as we saw, the automorphism part is not relevant when searching for maximal $P$-decompositions since it does not change the degree of a decomposition). 
By definition of $\mathcal{G}_P$, the following two operations over a generator matrix $G$ will provide matrices generating equivalent codes to the one generated by $G=(g_{ij})$:
\begin{enumerate}
	
	\item[(OP 1)] If $g_{i_0 j_0}\neq0$, $g_{ij_0}=0$ for every $i\neq i_0$ ($g_{i_0 j_0}$ is
	the only non-zero entry on the $j_0$-th column) and $r\leqslant_P j_0$ ($r\neq j_0$), we may assume $g_{i_0 r}=0$;
	
	This is equivalent to choosing the isometry $T\in \mathcal{G}_P$ such that $T(e_j)=e_j$ for every $j \neq j_0$ and 
	\[
	T(e_{j_0})=e_{j_0}-g_{i_0 r}g_{i_0 j_0}^{-1} e_r.	
	\]
	\item[(OP 2)] More generally, if there is $s\in[n]$ such that $r\leqslant_P j_i$ and $r\neq j_i$ for every $i\in\{1,\ldots,s\}$, furthermore, there are two rows of $G$, namely, $i_1,i_2\in [k]$, such that $g_{i_1r}=\sum_{l=1}^{s}x_l g_{i_1 j_l}$ and $g_{i_2r}=\sum_{l=1}^{s}x_l g_{i_2 j_l}$ for some choice of $x_1,\ldots,x_l\in\mathbb{F}_q$, and $g_{ij_l}=0$ for every $i\neq i_1,i_2$, 
	we may assume $g_{i_1r}=g_{i_2r}=0$. The procedure
	can be performed simultaneously to many lines and all those entries may be considered to be $0$. Furthermore, if the column $r$ is a linear combination of columns $j_{1},\ldots,j_{s}$, then one may exchange the column $r$ by the null column. Let $\{g_1,\ldots,g_n\}$ be the set of columns of $G$ and suppose
	\[
	g_{r}=\sum_{i=1}^s x_i g_{j_i}.
	\]
	In order to exchange the $r$-th column of $G$ by a null column we  consider the isometry $T\in\mathcal{G}_P$ defined by $T(e_i)=e_i$ for every $i\not\in\{j_1,\ldots,j_s\}$ and 
	\[
	T(e_{j_i})=e_{j_i}-x_i e_{r}
	\]
	for every $i\in\{1,\ldots,s\}$.
\end{enumerate}

\begin{definition}\label{main}
	Let $G$ be a generator matrix of an $[n,k,\delta]_q$ code $\mathcal{C}$. If $G$ is in generalized reduced row echelon form  and none non-null entry of $G$ may be exchanged by zero using Operations (OP 1) or (OP 2), we say that $G$ is in $P$-\textit{canonical form}.
\end{definition}

\begin{example}
	Let $\mathcal{C}$ be the $[6,3]_2$ code with generator matrix given by  
	$$G=\left[
	\begin{array}{cccccc}
	0 & 0 & 1 & 1 & 0 & 1\\
	1 & 0 & 1 & 1 & 1 & 0\\
	1 & 1 &0 & 0 & 0 & 0 
	\end{array}\right].
	$$
	Consider the poset $P_1$ with order relations $1\leqslant_{P_1} 2$ and $3\leqslant_{P_1} 4$. Applying operations (OP  1) and (OP  2) we get the following matrix in $P_1$-canonical form:
	$$G^\prime=\left[
	\begin{array}{cccccc}
	0 & 0 & 0 & 1 & 0 & 1\\
	1 & 0 & 0 & 1 & 1 & 0\\
	0 & 1 &0 & 0 & 0 & 0 
	\end{array}\right].
	$$
	Furthermore, if we consider $P_2$ a poset such that $P_1\subset P_2$ and $4\leqslant_{P_2} 5$, then performing operation (OP 1) we get
	$$G^{\prime\prime}=\left[
	\begin{array}{cccccc}
	0 & 0 & 0 & 1 & 0 & 1\\
	1 & 0 & 0 & 0 & 1 & 0\\
	0 & 1 &0 & 0 & 0 & 0 
	\end{array}\right].
	$$
	It is clear that $G^{\prime\prime}$ is in $P_2$-canonical form. It is also clear that it determines a maximum $P_2$-decomposition for $\mathcal{C}$.
\end{example}


\begin{theorem}
	Let $G$ be a generator matrix of a code $\mathcal{C}$. If $G$ is in a $P$-canonical form then the decomposition determined by $G$ is a maximal $P$-decomposition of $\mathcal{C}$.
\end{theorem}

\begin{IEEEproof}
	Assume, without loss of generality, $\mathscr{C}=(\mathcal{C};\mathcal{C}_0; \mathcal{C})$. We will first show that $\mathscr{C}$ does not admit aggregations. Suppose there is a $P$-decomposition $\mathscr{C}^\prime=(T(\mathcal{C});\mathcal{C}_0^\prime; T(\mathcal{C}))$ where $T\in \mathcal{G}_P$ and 
	$\phi_T(supp(\mathcal{C}_0))\subset supp(\mathcal{C}_0^\prime)$. Since $T\in\mathcal{G}_P$, the map $\phi_T$ coincides with the identity map, hence $supp(\mathcal{C}_0)\subset supp(\mathcal{C}_0^\prime)$. 
	Let $G_2=T(G)$ be the matrix whose $i$-th row is obtained by the action of $T$ in the $i$-th row of $G=(g_{ij})$. Thus, $G_2$ is a generator matrix of $T(\mathcal{C})$. 
	If $i_0 \in supp(\mathcal{C}_0^\prime)$, then $i_0$ is a null column of $G_2$. The characterization of $\mathcal{G}_P$ ensures that for every row $g_i=(g_{i1},\ldots,g_{in})$ of $G$, the $i_0$-th coordinate of $T(g_i)$, which is null and denoted by $T(g_i)_{i_0}$, is given by
	\begin{equation}\label{tg}
	0= T(g_i)_{i_0} = \sum_{\genfrac{}{}{0pt}{}{j}{i_0\leqslant_P j}} x_{ji_0}g_{ij} =g_{ii_0}x_{i_0i_0}+\sum_{\genfrac{}{}{0pt}{}{j}{j\neq i_0,\ i_0\leqslant_P j}} x_{ji_0}g_{ij}
	\end{equation}  
	for every $i\in\{1,\ldots,k\}$. Since $x_{i_0i_0}\neq 0$, it follows that  
	\begin{equation}\label{eeq01}
	g_{ii_0}=\sum_{\genfrac{}{}{0pt}{}{j}{j\neq i_0,\ i_0\leqslant_P j}} \left(-\frac{x_{ji_0}}{x_{i_0i_0}}\right)g_{ij}
	\end{equation}
	for every $i\in\{1,\ldots,k\}$. Since $G$ is in $P$-canonical form, Equation \ref{eeq01} together with Operation (OP 2) ensures that the $i_0$-th column of $G$ is null, so $i_0\in supp(\mathcal{C}_0)$. Therefore, $supp(\mathcal{C}_0^\prime)= supp(\mathcal{C}_0)$.
	
	The decomposition $\mathscr{C}$ also does not admit a splitting, indeed, suppose otherwise and let $\mathscr{C}^\prime= (T(\mathcal{C});\mathcal{C}_0;\mathcal{C}_i)_{i=1}^2$ be its refinement. 
	Let $G_1$ be as in Proposition \ref{maxrow}, then $G_1$ is a generator matrix of $T(\mathcal{C})$ and, by construction, the decomposition determined by $G_1$ is $\mathscr{C}^\prime$. Let $\beta_1$ be the minimal set of rows of $G$ such that for every $i\in\{1,\ldots,k_1\}$,
	\begin{equation}\label{01eq}
	w_i=\sum_{g\in\beta_1} x_g^i T(g)
	\end{equation}
	where $x_v^i\in\mathbb{F}_q$. Let $\beta_2$ be the minimal set of rows of $G$ such that
	\begin{equation}\label{02eq}
	w_i=\sum_{g\in\beta_2} x_g^i T(g)
	\end{equation}
	for every $i\in \{k_1+1,\ldots,k\}$. We stress that the only difference between this equalities and the ones in Proposition \ref{maxrow} is the presence of the linear isometry $T$. As before, we also have that $\beta_1\cap \beta_2 = \emptyset$. By construction, $supp(\beta_1)\cap supp(\beta_2)\neq \emptyset$. Let $i_0$ be the rightmost column of $G$ such that $i_0\in supp(\beta_1)\cap supp(\beta_2)$, if $i_0\not\in supp(T(\beta_1))$ then for every $g_i=(g_{i1},\ldots,g_{in})\in\beta_1$ (which is also a row of $G$), the $i_0$-th coordinate of $T(g_i)$ is given by Equation \ref{tg}.
	Since $i_0$ is the rightmost column of $G$ such that $i_0\in supp(\beta_1)\cap supp(\beta_2)$. Equation \ref{eeq01} ensures that the $i_0$-th column of $G$ is null, a contradiction. 
	The same argument shows that $i_0\in supp(T(\beta_2))$. Therefore, $i_0\in supp(T(\beta_1))\cap supp(T(\beta_2))$. By identities \ref{01eq} and \ref{02eq}, there exist $j\in \{1,\ldots,k_1\}$ and $l\in\{k_1+1,\ldots,k\}$ such that $i_0\in supp(w_j)\cap supp(w_l)$, a contradiction. 
\end{IEEEproof}


Each time we exchange a non-null column by a null column, we exchange a code $\mathcal{C}$
with $P$-decomposition $\mathscr{C}=\left(\mathcal{C};\mathcal{C}_{0};\mathcal{C}_{i}\right)_{i=1}^{r}  $ by a $P$-equivalent code with $P$-decomposition $\mathscr{C}%
^{\prime}=(  \mathcal{C}^{\prime};\mathcal{C}_{0}^{\prime};%
\mathcal{C}_{i}^{\prime})_{i=1}^{r}  $ where $dim (\mathcal{C}_{0}^{\prime})=dim (\mathcal{C}_{0})+1$. If the Operation (OP 2) is performed in a proper subset of $k$-lines of $G$, we do not increase
$dim (\mathcal{C}_{0})$ but, we may split some of the $\mathcal{C}_{i}$ into $\mathcal{C}_{i}=\mathcal{C}_{i_{1}%
}^{\prime}\oplus \mathcal{C}_{i_{2}}^{\prime}$ with $supp(\mathcal{C}_{i_{1}%
}^{\prime})  \cap supp(  \mathcal{C}_{i_{2}}^{\prime})
=\emptyset$.

Note that the role of $P$ in such operations rests solely on the
condition  $j\leqslant_P j_{1},j_{2},\ldots,j_s$. For the two extremal posets,
namely, the anti-chain and the chain poset, the picture is absolutely clear: If
$P$ is an anti-chain (hence we are considering the Hamming metric), no such
operation may be performed (since $i\leqslant_P j\iff i=j$). Hence, maximal $P$-decompositions coincide with maximal decompositions (see the paragraph after Definition \ref{com}) and the $P$-canonical form is a permutation of the classical systematic form; if $P$ is a chain
with $1\leqslant_P 2\leqslant_P \cdots\leqslant_P n$, then (OP 1) may be
performed to every $j\left(  i\right)  $ in the reduced row echelon matrix
$G$, hence, $\mathcal{C}$ is equivalent to a code that has a generator
matrix $G=(  g_{ij})  $ where $g_{ij (  i )  }=1$ and
$g_{ij}=0$ if $j\neq j (  i)  $ (see 
\cite{panek2010classification}). More general, for $NRT$ metrics, the $P$-canonical form is also a permutation of the $NRT$-triangular form presented in \cite{alves2011standard}. We stress that, fixing the natural labeling in the posets, the permutations necessary to obtain one form from other are the ones induced by $\mathcal{A}ut(P)$. 

The other case that can be easily described is the case of hierarchical posets. The algorithm to find $P$-decompositions according to the levels of the poset was first proposed in \cite{felix2012canonical}. 
The matrix obtained by this form is called canonical-systematic and it is also a permutation of the $P$-canonical form. Furthermore, in this case, the canonical-systematic form is a standard representation in the sense that every code has such decomposition. In \cite{MachadoPinheiroFirer}, it was proved that for every non-hierarchical poset, it is not possible to give a standard representation like this one (whose support of each component is entirely contained in a specific level of the poset). 

\section{Comparison of Posets and Decompositions}\label{sec:refinement}


As proved in \cite[Corollary 1]{felix2012canonical}, if $P$ is a hierarchical poset, a linear code may be canonically decomposed as $T(\mathcal{C})=\mathcal{C}_1\oplus \cdots\oplus \mathcal{C}_r$ where $supp\left(\mathcal{C}_{i}\right)\subset \Gamma_P^i$. As we shall explain in Section \ref{sec:applications}, this canonical decomposition permits to give explicit formulae and constructions for metric invariants. Since this decomposition (according to the levels of the poset) is a characteristic of hierarchical posets, see \cite[Theorem 3]{MachadoPinheiroFirer}, we aim to use the knowledge about hierarchical poset metrics to establish bounds in the general case. To do that, we need to show that the  $P$-primary decomposition is \textquotedblleft  well behaved\textquotedblright, in the sense that the decomposition are comparable when the posets are comparable. The meaning of the expression ``well behaved'' will be explained in Theorem \ref{well}.

\subsection{Decompositions and refinements}\label{sub:decomp_refine}

The set of all posets over $[n]$ is denoted by $\mathcal{P}_n$. It is, on itself, a partially ordered set with the order relation given by $P\leq Q$ if for every $a,b\in[n]$ with $a\leqslant_P b$ implies $a\leqslant_Q b$.

The set $\mathcal{P}_n$ has $n!$ (isomorphic) \emph{chain} posets and one \emph{anti-chain} poset, these $n!+1$ elements may be considered as extremal posets, indeed, it is easy to verify that adding a relation to a chain will give rise to a cycle $a\leqslant b\leqslant a$ since a total order on $[n] $ is defined once we state that $\sigma (1)\leq \sigma (2)\leq\ldots\leq\sigma (n)$ for some permutation $\sigma\in \mathcal{S}_n$, contradicting the anti-symmetry property of a partial order. On the other hand, it is also immediate to realize that no relation can be removed from an anti-chain since $a\leqslant a$ for all $a\in [n]$ are the unique relations on an anti-chain.


\begin{theorem}\label{well}
	\label{conj2}Let $P,Q\in\mathcal{P}_n$ with
	$P\leq Q$. Given a code $\mathcal{C}$, there is a maximal $P$-decomposition of $\mathcal{C}$ which is a $Q$-decomposition of $\mathcal{C}$.
\end{theorem}

\begin{IEEEproof}
	Assume $P,Q\in\mathcal{P}_n  $ and $P<Q$. Let
	$\mathscr{C}^{\prime}=\left(  \mathcal{C}^{\prime};\mathcal{C}_{0};\mathcal{C}_{i}\right)_{i=1}^r  $ be a maximal $P$-decomposition
	of $\mathcal{C}$ and $T\in GL_{P}(\mathbb{F}_q^n)$ such that $T\left(
	\mathcal{C}\right)  =\mathcal{C}^{\prime}$. By the characterization of $GL_P(\mathbb{F}_q^n)$, $T=A\circ T_\phi$ where $A\in\mathcal{G}_P$ and $T_\phi\in\mathcal{A}ut(P)$. From Proposition \ref{lema1}, we have that
	$$
	\mathscr{C}^{\prime\prime} = \left(  T_{\phi^{-1}}\left(  \mathcal{C}^{\prime}\right)  ;T_{\phi^{-1}}\left(
	\mathcal{C}_{0}\right)  ;T_{\phi^{-1}}\left(  \mathcal{C}_{i}\right)  \right)_{i=1}^r  
	$$
	is also a maximal $P$-decomposition of $\mathcal{C}$. However $T_{\phi^{-1}}\left(  \mathcal{C}^{\prime}\right)=T_{\phi^{-1}}\circ T (\mathcal{C})$, hence 
	$$
	T_{\phi^{-1}}\left(  \mathcal{C}^{\prime}\right)   =T_{\phi^{-1}}\circ A \circ T_\phi ( \mathcal{C}).
	$$
	We stress that $\mathcal{G}_P\subset \mathcal{G}_Q$ always that $P\leq Q$. Because $\mathcal{G}_P$ is a normal subgroup of $GL_P(\mathbb{F}_q^n)$, it follows that $T_{\phi^{-1}}\circ A \circ T_\phi \in \mathcal{G}_P$, hence $T_{\phi^{-1}}\circ A \circ T_\phi \in \mathcal{G}_Q$. Therefore, $\mathscr{C}^{\prime\prime}$ is a $Q$-decomposition of $\mathcal{C}$.
\end{IEEEproof}
 As a direct consequence of the previous theorem, we obtain a relation among primary decompositions of posets and the natural order over $\mathcal{P}_n$. 
 \begin{corollary}\label{col10}
	 If $P,Q\in\mathcal{P}_n  $ with
	 $P\leq Q$. Then, 
	 $\mathcal{D}_P(\mathcal{C})\leq \mathcal{D}_Q(\mathcal{C})$
	 for every linear code $\mathcal{C}$.
 \end{corollary}

Concerning primary $P$-decompositions, there is always a code that it is differently decomposed depending on the poset, that is, the preceding inequality is strict:

\begin{proposition}
	\label{conj3}If $P,Q\in\mathcal{P}_n$
	with $P<Q$. Then, there is a code $\mathcal{C}$ such that 
	$$\mathcal{D}_P(\mathcal{C})<\mathcal{D}_Q(\mathcal{C}).$$	
\end{proposition}

\begin{IEEEproof}
Suppose $P< Q$. Hence, there are $i_0,j_0 \in [n]$ such that 
	$$
	i_0   \nleqslant_{P} j_0 \text{ and }\ i_0 \leqslant_{Q} j_0.
	$$
	Let $\mathcal{C}=span\{e_{i_0}+e_{j_0}\}$ and $\mathcal{C}^{\prime}=span\{e_{j_0}\}$ be two codes over $\mathbb{F}_q^n$. 
	The linear map $T$ defined by $T(e_j)=e_j$ for every $j \in [n]\setminus \{i_0,j_0\}$, $T(e_{i_0})=e_{i_0}$ and $T(e_{j_0})=e_{j_0}-e_{i_0}$, is a $Q$-linear isometry, by the characterization of $GL_Q(\mathbb{F}_q^n)$ given in Proposition \ref{propos1}. Therefore, the $Q$-decomposition 
$		\mathscr{C}^\prime=(\mathcal{C}^\prime; \mathcal{C}_0^\prime; \mathcal{C}^\prime)$
	of $\mathcal{C}$, where $\mathcal{C}_0^\prime=\{v\in\mathbb{F}_q^n \ : \ v_{j_0}=0\}$, is a primary $Q$-decomposition and it follows that $\mathcal{D}_{Q}(\mathcal{C})=1$.

	On the other hand, when considering the $P$-metric, the decomposition $\mathscr{C}=\left(  \mathcal{C};\mathcal{C}_{0};\mathcal{C} \right)$, where $\mathcal{C}_0=\{v\in\mathbb{F}_q^n \ : \ v_{i_0}\neq 0 \text{ and } v_{j_0}\neq 0\}$, can not be refined since $\mathcal{C}$ is a one-dimensional code (hence, splittings are not allowed) whose support is composed by two maximal elements of $P$ (hence, aggregations are not allowed). Therefore, $\mathscr{C}^\prime$ is a primary $P$-decomposition for $\mathcal{C}$ and $\mathcal{D}_P(\mathcal{C})=0$.

\end{IEEEproof}

We remark that Corollary \ref{col10} together with Proposition \ref{conj3} implies that primary decomposition is a characterization of posets, in the sense that a given poset $P$ may be reconstructed from the profile of codes according to $P$. Moreover, looking at the proof of Proposition \ref{conj3}, one may notice that the reconstruction can be done by considering only the $n(n-1)/2$ pairs of vectors $(e_i,e_j)$.

\subsection{Hierarchical Bounds}\label{sub:hierar_bouds}

As we have mentioned, metrics induced by hierarchical posets are well understood. This knowledge was mainly obtained from the existence of the decomposition derived from the canonical-systematic form. Since the degree of a decomposition express the amount of $1$-step refinements performed from the trivial decomposition, we aim to use the knowledge of hierarchical posets and establish bounds for the degree considering the easy-to-compute degree of primary $P$-decompositions relatively to hierarchical posets.

As we have mentioned, hierarchical poset metrics are well understood and, in particular, if $P$ is a hierarchical poset, the profile (and consequently the degree) of a primary $P$-decomposition of a code $\mathcal{C}$ 
can be easily computed once we know the canonical-systematic form. 
For this reason, when considering a general poset $P$, we aim to establish bounds for $\mathcal{D}_P(\mathcal{C})$ considering the easy-to-compute primary $P$-decompositions relatively to hierarchical posets.

Considering the natural order $\leq$ on $\mathcal{P}_{n}$, out of $P$ we can define two hierarchical posets:

 \textbf{1 - Upper neighbor:} 
$$P^{+}  =\min\left\{  Q\in\mathcal{P}_{n} \ : \ P\leq Q \text{ and } Q\text{ is hierarchical}%
 \right\}.$$

 \textbf{2 - Lower neighbor: }
$$P^{-} =\max\left\{  Q\in\mathcal{P}_{n} \ : \ Q\leq P \text{ and } Q\text{ is hierarchical}\right\}.$$

The next proposition follows directly from Corollary \ref{col10}.

\begin{proposition}\label{bound} For any linear code $\mathcal{C}$, 
\[
\mathcal{D}_{P^{+}}\left(
\mathcal{C}\right)  \leq \mathcal{D}_{P}\left(  \mathcal{C}\right) \leq  \mathcal{D}_{P^{-}}\left(  \mathcal{C}\right).
\]

\end{proposition}

If $P$ is not hierarchical, both inequalities are strict for some code
$\mathcal{C}$. Moreover, the bounds are tight, in the sense that, given a poset $P$, there are codes $\mathcal{C}_1$ and $\mathcal{C}_2$ such that $\mathcal{D}_{P^{+}}\left(
\mathcal{C}_1\right)  =\mathcal{D}_{P}\left(  \mathcal{C}_1\right)  $ and  $\mathcal{D}_{P^{-}%
}\left(  \mathcal{C}_2\right)  =\mathcal{D}_{P}\left(  \mathcal{C}_2\right)  $ (just consider any code $\mathcal{C}$ with $supp(\mathcal{C})\subset \Gamma_P^1$).

\medskip 

\section{Applications to Coding Theory}\label{sec:applications}

The poset-metrics, as much as the decomposition problem, arose in the context of coding theory, what justified the use of the terminology and the notation from the area: code and codeword for vector subspace and vector, respectively. Now comes the time to justify it with some content, i.e., with applications to coding theory.

 We combine the knowledge available for hierarchical posets with the degree's bound found in Section \ref{sub:hierar_bouds}  to produce bounds for the most important invariant in coding theory (the packing radius) and to optimize the traditional full lookup table searching algorithm in the syndrome decoding process.

\subsection{Packing Radius Bounds}\label{sub:pack}

	If $\mathbb{F}_q^n$ is endowed with a $P$-metric, an $[n,k,\delta_P]$ $P$-code is an $[n,k]$-linear code $\mathcal{C}$ with \emph{minimum distance}
	\[
	\delta_P=\min \{w_P(c) : c\in\mathcal{C}\} = \min \{ d_P(c,c'): c,c'\in\mathcal{C},c\neq c'  \}.
	\]  
	In the context of coding theory, the minimum distance is considered the most important metric parameter of a code. Its importance comes from the fact that it may ensures the correction of a certain amount of errors. If $P$  is an anti-chain, the $P$-weight and metric are known as Hamming weight and metric, denoted by $w_H$ and $\delta_H$, respectively. This is the most important setting in the context of coding theory. In this situation, let us say that a message $c\in \mathcal{C}$ is transmitted  and it is received with errors as a message $v=c+e$, where $e\in\mathbb{F}_q^n$ is the error vector. If $\delta_H$ is the minimum distance of $\mathcal{C}$ and $w_H(e)\leq \lfloor (\delta_H-1)/2 \rfloor$ (where $ \lfloor \cdot \rfloor  $ is the floor function), then we can ensure that the original message $c$ can be recovered. The quantity $\lfloor (\delta_H-1)/2 \rfloor$ happens to be what is known as the \emph{packing radius} of the code, defined, in general, as 
	\[
	\mathcal{R}_P(\mathcal{C})=\max\{r\geq 0: B_P(c,r)\cap B_P(c',r)=\emptyset \ \forall c,c'\in\mathcal{C},c\neq c'  \},
	\] 
	where $B_P(u,r)=\{v\in\mathbb{F}_q^n:d_P(u,v)\leq r   \}$ is the \emph{$P$-ball} with center $u$ and radius $r$.
	
	When considering a general poset metric, the packing radius maintains the same property of evaluating the correction capability. However, in general, the minimum distance does not determine the packing radius, only bounds for it (see \cite{d2015packing}). Indeed, hierarchical posets are the unique posets for which the packing radius of a code is a function of its minimum distance, as can be seen in \cite{MachadoPinheiroFirer}. Determining the packing radius, in general settings, is a very difficult task, it may be an NP-hard problem even considering the case of a code with only two elements (see \cite{d2015packing}). For this reason, working with a general poset, the best we can do is to find bounds and we do it considering maximal $P$-decompositions of the code. We start with two simple propositions.


\begin{proposition}\label{prop15}
	Let $\mathscr{C}=(\mathcal{C}^\prime;\mathcal{C}_0;\mathcal{C}_i)_{i=1}^r$ be a $P$-decomposition for $\mathcal{C}$. Then,
	\[
		\mathcal{R}_P(\mathcal{C}) \leq \min_{i\in\{1,\ldots,r\}} \mathcal{R}_P(\mathcal{C}_i).
	\]
\end{proposition}

\begin{IEEEproof}
	Note that $\mathcal{R}_P(\mathcal{C})=\mathcal{R}_P(\mathcal{C}^\prime)$. Furthermore, since each $\mathcal{C}_i$ is a subcode of $\mathcal{C}^\prime$, it follows that $\mathcal{R}_P(\mathcal{C})\leq \mathcal{R}_P(\mathcal{C}_i)$ for every $i\in \{1,\ldots,r\}$.
\end{IEEEproof}


\begin{proposition}\label{bound01}
	If $P\leq Q$, then $ \mathcal{R}_P(\mathcal{C})\leq  \mathcal{R}_Q(\mathcal{C})$ for every linear code $\mathcal{C}$.
\end{proposition}

\begin{IEEEproof}
	If follows directly from the fact that 
	$$B_Q(r,c)\cap B_Q(r,\mathbf{0})\subset B_P(r,c)\cap B_P(r,\mathbf{0})$$ for every $c\in\mathcal{C}$ and any integer $r\geq 0$. 
\end{IEEEproof}

Using Proposition \ref{bound01} and the upper and lower neighbours $P^+$ and $P^-$ defined in the previous section, bounds for the packing radius of codes according to hierarchical posets may be obtained. The proof of the next proposition follows straight from the previous proposition.
\begin{proposition}\label{prop16}
	Given a maximal $P$-decomposition $\mathscr{C}=(\mathcal{C}^\prime;\mathcal{C}_0;\mathcal{C}_i)_{i=1}^r$ for $\mathcal{C}$, 
	\[
		\mathcal{R}_{P^-}(\mathcal{C}_i)\leq \mathcal{R}_P(\mathcal{C}_i) \leq \mathcal{R}_{P^+}(\mathcal{C}_i)
	\]
	for every $i\in\{1,\ldots,r\}$.
\end{proposition}

Propositions \ref{prop15} and \ref{prop16} yield the following upper and lower bounds for the packing radius:
\begin{equation}\label{eq11}
	\mathcal{R}_{P^-}(\mathcal{C}) \leq \mathcal{R}_P(\mathcal{C})\leq \min_{i\in\{1,\ldots,r\}}\mathcal{R}_{P^+}(\mathcal{C}_i)
\end{equation}
for every $i\in\{1,\ldots,r\}$.

Since $P^+$ and $P^-$ are hierarchical, these bounds are obtained just by finding the minimum distance of $\mathcal{C}$ and each code $\mathcal{C}_i$ according to $P^-$ and $P^+$, respectively. Furthermore, each of these codes may be decomposed according to the lower and upper neighbors in order to simplify and optimize the search for the minimum distance. If $P$ is hierarchical, we obtain equalities in the bounds obtained in Proposition \ref{prop16} and in Inequality (\ref{eq11}), hence, these bounds are tight.

\subsection{Optimizing the Full Lookup Table Searching in Syndrome Decoding}\label{sub:syndrome}


The core of coding theory is the existence of errors: the messages to be sent are elements of a code $ \mathcal{C}\subset\mathbb{F}_q^n $, but, in the occurrence of errors, any element of $\mathbb{F}_q^n $ may be received. Once $y\in \mathbb{F}_q^n $ arrives, the receiver needs to decide what was the original message. There are essentially to types of decision criteria: probabilistic and deterministic. We are interested in the deterministic criteria determined by metrics, the well-known \textit{minimum distance decoding}.

Minimum distance decoding works as follows: once $y\in \mathbb{F}_q^n $ is received, we look for one of the codewords that minimizes the distance to $y$. We are in a good situation if the minimum is unique and this is the case if the weight of the error was not greater than the packing radius. 
Despite the fact that the minimum distance decoding is very simple to describe as a criterion, its implementation as an algorithm is very difficult, since codes (for practical purposes) are very large and searching on very large sets is infeasible. 

Syndrome decoding is an algorithm to perform minimum distance decoding that works for every linear code and every metric that is determined by a weight or, equivalently,  a metric that is invariant by translations. 
Given a code $ \mathcal{C}\subset\mathbb{F}_q^n $, for each coset $u+ \mathcal{C}$, consider an element with minimal weight, called coset leader. In syndrome decoding, once a message $y$ is received, instead looking for $c\in\mathcal{C}$ that minimizes $d_P(y,c)$, one looks for a coset leader $u$ such that $T(u-y)=0$, where $T:\mathbb{F}_q^n\rightarrow \mathbb{F}_q^{n-k}$ is a linear map that has $\mathcal{C}$ as its kernel (determined by the parity check matrix, in coding jargon). The advantage of using syndrome decoding is the exchange of a lookup table of size $q^k$ by one of size $q^{n-k}$, what is smaller, in general, since coding theory looks for $[n,k]$-codes with $k/n$ close to $1$. This quantity, $q^{n-k}$, is an algebraic invariant that does not depend on the metric.

We do not explain here the details of syndrome decoding (it can be found in any book on coding theory, for example \cite[Section 1.11]{huffman2003fundamentals}), but it is a very established process and all we need is to state that when we use the poset metric $d_P$, we mean that the minimality of the coset leader refers to the $P$-weight $w_P$. 

In this section we show how we can reduce the searching process in the lookup table of syndrome decoding, by  considering $P$-decompositions of a code. Roughly speaking, the finest the decomposition, the smaller the search table. Let us do it in details.



Let $\mathscr{C}=\left(  \mathcal{C};\mathcal{C}_{0};\mathcal{C}_{i}\right)_{i=1}^r$ be a maximal $P$-decomposition of an $[n,k,\delta]_q$ code $\mathcal{C}$. 
Initially, note that in order to perform decoding, we can ignore the component $V_{0}$ (recall that $V_i$ is the support-space of $\mathcal{C}_i$) since we know that any codeword $c=(c_1,\ldots,c_n)\in \mathcal{C}$ should have $c_{j}=0$ for every $j\in\left[n\right]  ^{\mathcal{C}}$. Consider the projection 
$\pi:\mathbb{F}_q^n \longrightarrow \mathbb{F}_q^{n_1+\cdots+n_r}$ (recall that $n_i=|supp(\mathcal{C}_i)|$) defined by
\[
\pi(x_1,\ldots,x_n) = (x_{i_1},\ldots,x_{i_s})
\]
where  $\{i_1,\ldots,i_s\}=supp(\mathcal{C})$ and we assume $i_1<\cdots<i_s$ . The map $\pi_{1,\ldots,r}=\pi|_{\oplus_{i=1}^r V_i}$, the restriction of $\pi$ to the space $\oplus_{i=1}^r V_i$, is a bijection.  Therefore, by pushing forward the metric in the restriction, we obtain the metric $d_P^\pi$ in $\mathbb{F}_q^{n_1+\cdots+n_r}$ defined by
\[
d_P^\pi(x,y):=d_P(\pi_{i,\ldots,r}^{-1}(x),\pi_{i,\ldots,r}^{-1}(y))
\]
for every $x,y\in\mathbb{F}_q^{n_1+\cdots+n_r}$. The metric $d_P^\pi$ turns the restriction map $\pi_{1\ldots r}$ into a linear isometry. Because $\mathcal{C}$ is a subspace of $V_1\oplus \ldots\oplus V_r$, the proof of the proposition below follows straight from these observations. 
\begin{proposition}\label{propo}
	The metric-decoding criteria of $\mathbb{F}_q^n$ for $\mathcal{C}$ is equivalent to the metric-decoding criteria of $\mathbb{F}_q^{n_1+\cdots+n_r}$ for $\pi(\mathcal{C})$, i.e.,
	\[
	d_P(c^\prime,y)=\min_{c\in \mathcal{C}} d_P(c,y) 
	\iff 
	d_P^\pi(\pi(c^\prime),\pi(y))=\min_{c\in \pi(\mathcal{C})} d_P^\pi(c,\pi(y)).
	\]
\end{proposition}

By Proposition \ref{propo}, to perform syndrome decoding, instead of using a lookup table with $|\mathbb{F}_q^n/\mathcal{C}|=q^{n-k}$ elements, we can reduce the number of cosets to $|\mathbb{F}_q^{n_1+\cdots+n_r}/\pi(\mathcal{C})|=\prod_{i=1}^r q^{n_i-k_i}$ elements. Note that 
$$q^{n_0} \times \prod_{i=1}^r q^{n_i-k_i} = q^{n-k}.$$
Therefore, $P$-decompositions having the maximum possible number of elements in the support of $\mathcal{C}_0$ are the best ones in order to perform syndrome decoding. 


Besides this possible (and a-posteriori irrelevant) gain in the cardinality of the syndrome lookup table obtained considering aggregations, there is a more significant gain that can be obtained through the splitting operation.

\begin{proposition}\label{prop001}
	If 
	$$\langle supp(\mathcal{C}_i)\rangle_P \cap \langle supp(\mathcal{C}_j)\rangle_P = \emptyset $$
	for all $i\neq j$ and $i,j\neq 0$, then given $y\in\mathbb{F}_q^n$, 
	\[
	\min_{c\in \mathcal{C}} d_P(c,y) = \sum_{i=1}^r \min_{c\in \mathcal{C}_i} d_{P}^{\pi_i}(\pi_i(c),\pi_i(y)).
	\]
\end{proposition}

\begin{IEEEproof}
	Note that if $c\in\mathcal{C}$, then $c=c_1+\cdots+c_r$ with $c_i\in\mathcal{C}_i$ and for every $y_i\in\mathbb{F}_q^{n_i}$,
	\[
	supp(y_i-c_i)\subset \langle supp(\mathcal{C}_i)\rangle=supp(V_i).
	\] 
	Then,
	\[
	d(y,c)=d(y_1,c_1)+\cdots+d(y_r,c_r),
	\]
	where $y=y_1+\cdots+y_r$ and $y_i\in\mathbb{F}_q^{n_i}$ for all $i\in [r]$. 
\end{IEEEproof}

Due to Proposition \ref{prop001}, if the ideals generated by each component are disjoint, syndrome decoding can be done independently in each component $\mathcal{C}_i$. Therefore, the number of cosets can be reduced from $q^{n-k}$ elements to $\sum_{i=1}^r q^{n_i-k_i}$ elements, where the last one is the sum of the cosets in each quotient $\mathbb{F}_q^{n_i}/\pi_i(\mathcal{C}_i^{\prime})$. More generally, if $r$ is a disjoint union of subsets $I_i$, i.e., $[r]=I_1 \sqcup \ldots \sqcup I_s$, and 
\[
\langle supp(\oplus_{i\in I_j} \mathcal{C}_i)\rangle_P \cap \langle supp(\oplus_{i\in I_l} \mathcal{C}_i) \rangle_P=\emptyset,
\]
for every $j\neq l$, then decoding can be separately done in each projection of $\oplus_{i\in I_j} \mathcal{C}_i$ into $\mathbb{F}_q^{N}$ where $N=\sum_{i\in I_j} n_i$.

Until now, in order to obtain the reductions, it was not necessary to change the syndrome decoding algorithm; we just performed it in a different (smaller) space. The hierarchical relation among elements of the poset will provide us a ``quasi-independent'' syndrome decoding algorithm that is performed by choosing first coordinates that hierarchically dominate others, therefore we will call this algorithm a \index{Leveled Syndrome Decoding}\textit{Leveled Syndrome Decoding}. 

Given two subsets $I,J\subset [n]$, we say that $I$ and $J$ are \textit{hierarchically related} if every element in $I$ is smaller than every element in $J$. Suppose $[r]$ is an ordered disjoint union of sets, $[r]=I_1 \sqcup \ldots \sqcup I_s$, such that $supp(\oplus_{i\in I_j} \mathcal{C}_i)$ is hierarchically related with $supp(\oplus_{i\in I_{j+1}}\mathcal{C}_i)$ for every $j\in\{1,\ldots,s-1\}$, so if $i_0\in supp(\oplus_{i\in I_{j_0}} \mathcal{C}_i)$, then $i_0\leqslant i$ for every $i\in supp(\oplus_{i\in I_l} \mathcal{C}_i)$ with $l>j_0$.
By using the well-known syndrome decoding algorithm, the leveled syndrome decoding algorithm is as follows: 
\bigskip

\textbf{Input:} $y=y_1+\cdots+y_s \in\mathbb{F}_q^n$ where $y_i\in \oplus_{j\in I_i} V_j$

\hspace{0.6cm}For each $i\in\{1,\ldots,s\}$ do

\hspace{0.9cm}Decode $\pi_{j\in I_i}(y_i)\in\mathbb{F}_q^{\sum_{j\in I_i} n_j}$ using syndrome ($d_P^{\pi_{j\in I_i}}$) outputting $c_i\in \pi_{j\in I_i}(\oplus_{j\in I_i} \mathcal{C}_j)$

\textbf{Output:} $c=\pi_{j\in I_1}^{-1}(c_1)+\cdots+\pi_{j\in I_s}^{-1}(c_s)$.
\begin{center}
	\hspace{0.3cm}Leveled Syndrome Decoding Algorithm 1.	
\end{center}




Furthermore, performing an ordered syndrome decoding from the higher level to the smaller ones, we can proceed as follows:

\bigskip

\textbf{Input:} $y=y_1+\cdots+y_s \in\mathbb{F}_q^n$ where $y_i\in \oplus_{j\in I_i} V_j$

\hspace{0.3cm}For $i=s$ to $1$, do

\hspace{0.6cm}If $\pi_{j\in I_i}(y_i)\in \oplus_{j\in I_i} \mathcal{C}_j$ do

\hspace{0.9cm}$c_i=\pi_{j\in I_i}(y_i)$;

\hspace{0.6cm}else do

\hspace{0.9cm}Decode $\pi_{j\in I_i}(y_i)\in\mathbb{F}_q^{\sum_{j\in I_i} n_j}$ using syndrome ($d_P^{\pi_{j\in I_i}}$) outputting $c_i\in \pi_{j\in I_i}(\oplus_{j\in I_i} \mathcal{C}_j)$;

\hspace{0.9cm}Go to Output;

\hspace{0.6cm}end if;

\hspace{0.3cm}end For;

\textbf{Output:} $c=\pi_{j\in I_i}^{-1}(c_i)+\pi_{j\in I_{i+1}}^{-1}(c_{i+1})+\cdots+\pi_{j\in I_s}^{-1}(c_s)$.

\begin{center}
	\hspace{0.3cm}Leveled Syndrome Decoding Algorithm 2.	
\end{center}










The first algorithm was described in \cite{felix2012canonical} for the hierarchical poset case. The second one is also a minimum distance algorithm according to the poset metric $d_P$ (see the next proposition), but if there is an error in a particular level, the decoder outputs the null vector in the levels covered by the one where the error happened.  

\begin{proposition}
	Algorithm 2 determines a minimum distance decoder according to the metric $d_P$.
\end{proposition}

\begin{IEEEproof}
	Given $y\in\mathbb{F}_q^n$ and $c\in\mathcal{C}$, then
	\[
	d_P(y,c)=d_P^{\pi_{j\in I_i}}(\pi_{j\in I_i}(y),\pi_{j\in I_i}(c))
	\] 
	where $i$ is the largest integer such that $\pi_{j\in I_i}(y)\neq \pi_{j\in I_i}(c)$. Therefore, if $c^\prime\in\mathcal{C}$ satisfies
	\[
	d_P(y,c^\prime)=\min_{c\in\mathcal{C}} d_P(y,c),
	\]
	then $c^{\prime\prime}=c^\prime_s+c^\prime_{s-1}+\cdots+c^\prime_i$ also attains the minimum and this is the codeword returned by algorithm 2.
\end{IEEEproof}

Both the algorithms demand the storage of the lookup tables of each quotient $\mathbb{F}_q^{N}/\pi_{j\in I_i}(\oplus_{j\in I_i} \mathcal{C}_j)$ where $\sum_{j\in I_i} n_j$. The total number of elements we need to store is
\begin{equation}\label{am1}
\sum_{i=1}^s \prod_{j\in I_i} q^{n_j-k_j}.
\end{equation}
The difference between Algorithm 1 and 2 is that while the search in the Algorithm 1 is performed over all elements of the lookup table, the ordered procedure in Algorithm 2 stop the search in the first level where an error has occurred, i.e., if this happens in the level $i_0$, then the search is performed only over the lookup table of $\pi_{j\in I_{i_0}}(\oplus_{j\in I_{i_0}} \mathcal{C}_j)$ which has
\begin{equation}\label{am2}
\prod_{j\in I_{i_0}} q^{n_j-k_j}
\end{equation}
elements. Note that if the poset has only one level, the values of Expressions (\ref{am1}) and (\ref{am2}) coincide and are equal to $q^{n-k}$. However, the more the support of a $P$-decomposition of a code is partitioned into hierarchically related subsets,  the more we may transform factors in the product  (\ref{am1}) into parcels in expression (\ref{am2}).


\section*{Acknowledgment}

The authors would like to thank the S\~{a}o Paulo Research Foundation (Fapesp) for the financial support through grants 2013/25977-7 and 2016/01551-9. The second author would like to thank also CNPq, for the support through grant 303985/2014-3.

\ifCLASSOPTIONcaptionsoff
  \newpage
\fi

    \bibliographystyle{plain} 
    \bibliography{biblio}




%
%
%





\end{document}